\documentclass[12pt]{article}
\pdfoutput=1
\usepackage{jheppub}

\usepackage{amsmath,amssymb,amsfonts,mathtools}
\usepackage{graphicx}
\usepackage{color}
\usepackage{tikz}
\usepackage{dsfont}
\usepackage[normalem]{ulem}
\usepackage{booktabs}
\usepackage{multirow}

\usepackage{hyperref}

\usepackage{subfigure}

\usepackage{comment}

\makeatletter\def\l@subsubsection#1#2{}%
\makeatother

\def\be{\begin{eqnarray}}
\def\ee{\end{eqnarray}}

\def\makeatletter{\catcode`\@=11}% 11:letter
\makeatletter
\def\mathbox#1{\hbox{$\m@th#1$}}%
\def\math@ccstyles#1#2#3#4#5#6#7{{\leavevmode
      \setbox0\mathbox{#6#7}%
      \setbox2\mathbox{#4#5}%
      \dimen@ #3%
      \baselineskip\z@\lineskiplimit#1\lineskip\z@
      \vbox{\ialign{##\crcr
             \hfil \kern #2\box2 \hfil\crcr
             \noalign{\kern\dimen@}%
             \hfil\box0\hfil\crcr}}}}
\def\mathaccstyles{\math@ccstyles\maxdimen}
\def\maththroughstyles{\math@ccstyles{-\maxdimen}}
\def\unity%
 {\maththroughstyles{.45\ht0}\z@\displaystyle {\mathchar"006C}\displaystyle 1}

\title{The SymTFT of $u(N)$ Yang-Mills Theory and Holography}

\author[a,b]{Oren Bergman,}
\author[c,d]{Eduardo Garcia-Valdecasas,}
\author[e]{Francesco Mignosa,}
\author[e]{Diego Rodriguez-Gomez}
\affiliation[a]{Department of Physics, University of California, San Diego \\[-3mm]}
\affiliation[b]{Department of Physics, Technion, Israel Institute of Technology, Haifa, 32000, Israel\\[-3mm]}
\affiliation[c]{SISSA, Via Bonomea 265, Trieste 34136, Italy\\[-3mm]}
\affiliation[d]{INFN - Sezione di Trieste, Via Valerio 2, 34127 Trieste, Italy\\[-3mm]}
\affiliation[e]{Department of Physics, Universidad de Oviedo, C/Federico Garcia Lorca 18, 33007 Oviedo, Spain}

\abstract{We propose a SymTFT for 4d $U(N)$ Yang-Mills theory and its variants. We show that the SymTFT 
reproduces the structure of the global one-form symmetry in these theories. We consider the holographic embedding 
of this SymTFT, and observe that SymTFT's containing continuous symmetries are not 
always obtained as a near boundary limit of the supergravity action. }

\begin{document}

\maketitle

\flushbottom

\section{Introduction}

It is well-appreciated now that extended operators play a central role in characterizing quantum field theories.
For example in gauge theories line operators differentiate between theories based on the same gauge algebra,
but with gauge groups that have a different global structure,
as well as different possible values of discrete theta-like parameters \cite{Aharony:2013hda}.
The different theories can also be characterized by the different finite one-form symmetries acting on the line operators \cite{Gaiotto:2014kfa}.
This idea has been recast and generalized by the concept
of the SymTFT, a topological quantum field theory in one higher dimension with two boundaries; one determining the local dynamics
of the theory, and the other determining the finite global symmetry and its anomalies \cite{Freed:2012bs,Apruzzi:2021nmk}.
For discrete abelian symmetries the SymTFT is generically a $U(1)$ $BF$ gauge theory. 
This construction has its origins in the AdS/CFT correspondence  \cite{Witten:1998wy}.
The global symmetries of the boundary CFT are dual to gauge fields in the bulk AdS, 
with appropriate boundary conditions. Roughly speaking, the SymTFT of the CFT  
is the near boundary limit of the supergravity action of these bulk gauge fields in AdS. 

For continuous symmetries the situation is less clear.
There is a recent proposal that the SymTFT of a theory with a global $U(1)$ symmetry is a $BF$ theory involving 
both a $U(1)$ gauge field and an $\mathbb{R}$
gauge field \cite{Brennan:2024fgj,Antinucci:2024zjp}. But it is not known in general how 
such a theory may arise in holography. Indeed, as is well known, a $U(1)$ global symmetry is dual to a massless $U(1)$ gauge field, which has a Maxwell action rather than a $BF$ action. 
A possible approach for relating the two was presented in \cite{Bonetti:2024cjk}.\footnote{For recent work see 
\cite{Gagliano:2024off, Yu:2024jtk}.}

In this paper we will study the global symmetry structure of the 4d $U(N)$ Yang-Mills theory and its variants.
These theories all have a $U(1)\times U(1)$ global one-form symmetry,
as well as a finite symmetry that depends on the variant.
Generalizing \cite{Brennan:2024fgj,Antinucci:2024zjp}, we will propose a 5d SymTFT for these theories, and show 
that it correctly reproduces the global symmetry structure.

We will also attempt to derive this SymTFT from holography for ${\cal N}=4$ supersymmetric Yang-Mills theory.
Our attempt will fail, indicating that the dual CFT is the $SU(N)$ theory and its variants.
More generally we will show that a SymTFT for a $d$-dimensional QFT with a $U(1)$ $p$-form symmetry cannot be embedded
in holography if $d \leq 2p + 3$.

The rest of the paper is organized as follows.
In section \ref{U(N)theories} we discuss the line operator spectrum and corresponding one-form symmetry of 4d 
$u(N)$ Yang-Mills theories.
We also explain how the different theories are related by gauging finite 
subgroups of the one-form symmetry. In section \ref{symTFTU(N)} we propose the 5d SymTFT for these theories, derive the properties of the topological operators, and reproduce the 
symmetry properties of the different 4d theories in terms of the possible boundary conditions. In section \ref{symTFTSU(N)} we discuss the connection between the $u(N)$
and the $su(N)$ SymTFTs via gauging $U(1)$.
In section \ref{holography} we address the question of holographic embedding, and derive a general condition for a SymTFT
of a theory with a $U(1)$ global symmetry to be realized in holography.
Section \ref{Conclusions} contains our conclusions.

\section{The $u(N)$ theories}\label{U(N)theories}

The $U(N)$ theory belongs to a family of gauge theories with an $su(N)\times u(1)$ gauge algebra,
with a particular choice of global structure of the gauge group. The general form of the gauge group is
$[SU(N)/\mathbb{Z}_k \times U(1)]/\mathbb{Z}_r$,
where $k$ is a divisor of $N$ and $r$ is a divisor of $N/k$. 
The group $\mathbb{Z}_r$ is embedded diagonally in the $\mathbb{Z}_{N/k}$ center of $SU(N)/\mathbb{Z}_k$ and the $U(1)$.
The $U(N)$ theory corresponds to the case $k=1$ and $r=N$.

\subsection{Spectrum and symmetry}

The different theories differ in their spectrum of line operators.
A generic line operator carries four charges $(z_e,z_m,n_e,n_m)$, with $z_e,z_m\in \mathbb{Z}_N$ and
$n_e,n_m \in \mathbb{R}$, and the spectrum is constrained by the Dirac pairing condition 
\be
\label{DiracPairing}
{z_e z_m' - z_m z_e'\over N} + n_e n_m' - n_m n_e'  \in \mathbb{Z} .
\ee
The spectrum of the $[SU(N)/\mathbb{Z}_k \times U(1)]/\mathbb{Z}_r$ theory is given by
\be
\label{LineOperators1}
(z_e,z_m,n_e,n_m) = \left(kn_1 + \ell n_2 , {N\over rk} n_2, n_1+ r n_3
, -{1\over r}(n_2-rn_4)\right) \, ,
\ee
where $n_1,n_2,n_3,n_4 \in \mathbb{Z}$, and where $\ell$ is a discrete theta parameter taking values in
$\{0,1,\ldots, rk-1\}$.
For $r=1$ we can shift $n_3, \, n_4$ such that the charges $(z_e,z_m)$ and $(n_e,n_m)$ are uncorrelated, 
and the former reduces to the
spectrum of the $[SU(N)/\mathbb{Z}_k]_{\ell}$ theory \cite{Aharony:2013hda}. 
For $r=N$, $k=1$ we obtain the spectrum of the $U(N)$ theory.\footnote{Note that the $U(N)$ theory admits a 
discrete theta parameter $\ell \in \{0,1,\dots N-1\}$. This is due to the fact that the $U(N)$ theory admits $SU(N)$ instantons with fractional charge $\in \mathbb{Z}/N$ (for example on $T^4$) \cite{Anber:2024uwl}.} 

 The global symmetry acting on the spectrum of line operators in (\ref{LineOperators1}) is given by
 \be
\label{GlobalSymmetry}
G[k,r,\ell] = U(1)^{(1)}_e \times U(1)^{(1)}_m \times \mathbb{Z}_{N/r gcd(k,N/rk,\ell)}^{(1)} \times \mathbb{Z}_{gcd(k,N/rk,\ell)}^{(1)} .
\ee
For $r=1$ this reduces to the global symmetry of the $[SU(N)/\mathbb{Z}_k]_\ell \times U(1)$ theory \cite{Gaiotto:2014kfa},
and for $k=1, r=N$ it reduces to just $U(1)^{(1)}_e \times U(1)^{(1)}_m$, which is the global symmetry of the $U(N)$ theory.
The finite part of the global symmetry is obtained as follows. The purely torsion charged lines have a spectrum 
$(z_e,z_m) = (rkn + r\ell m, Nm/k)$. This is acted on by the group
\be
\label{FiniteGlobalSymmetry}
 \left(\mathbb{Z}_{N/rk}^{(1)} \times \mathbb{Z}_{N/r gcd(N/kr,\ell)}^{(1)}\right)/\mathbb{Z}_{N/kr gcd(N/kr,\ell)}^{(1)} ,
\ee
where the first factor acts on the electric lines with $(z_e,z_m)=n(rk,0)$, the second factor acts on the dyonic lines
with $(z_e,z_m)=m(r\ell, N/k)$, and the quotient is due to the identification $\ell(rk,0) = k(r\ell, N/k)$.
One can show that this is isomorphic to the finite part of (\ref{GlobalSymmetry})
(see \cite{Bergman:2022otk} for the explicit transformation in the case with $r=1$).
There is a mixed anomaly between $U(1)^{(1)}_e$ and $U(1)^{(1)}_m$, and an order $\mbox{gcd}(k,N/rk,\ell)$ mixed anomaly between the finite symmetry factors.
This again reduces to the known properties of the $[SU(N)/\mathbb{Z}_k]_\ell$ theory when $r=1$ \cite{Gaiotto:2014kfa,Hsin:2020nts,Bergman:2022otk}.

\subsection{Finite gauging relations}

The different theories described in the previous section are related by gauging finite subgroups of the 
one-form global symmetry.
We will focus on the theories with a vanishing discrete theta parameter, $\ell  = 0$.
In this case the global symmetry is given by
\be
\label{GlobalSymmetry4}
G[k,r,0] = U(1)^{(1)}_e \times U(1)^{(1)}_m \times \mathbb{Z}_{N/rk}^{(1)} \times \mathbb{Z}_k^{(1)} .
\ee
Starting with the $U(N)$ theory ($k=1, r=N$), one can gauge 
any $\mathbb{Z}_{N_e}^{(1)} \times \mathbb{Z}_{N_m}^{(1)}$ subgroup of $U(1)^{(1)}_e \times U(1)^{(1)}_m$
consistent with the mixed anomaly. The anomaly action is given by
\be
S_{anomaly}[B_2,C_2] = {N\over 2\pi} \int_{M_5} dB_2 \wedge C_2\, ,
\ee
where $B_2,C_2$ are background fields for $U(1)_e^{(1)}$ and $U(1)_m^{(1)}$, respectively.\footnote{This is a straightforward
generalization of the anomaly action for the $U(1)$ theory \cite{Gaiotto:2014kfa}. Decompose the $U(N)$ gauge field into
its $SU(N)$ and $U(1)$ parts ${\cal A}_1 = a_1 + {1\over N} A_1 {\mathbb{I}}$. Turning on a background field $B_2$ for the electric symmetry
$U(1)^{(1)}_e$
corresponds to replacing $dA_1$ by $dA_1 + NB_2$ in the action. The factor $N$ is due to the fact that the 
%$A$
purely abelian Wilson line carries
$N$ units of charge. Turning on a background field $C_2$ for the magnetic symmetry $U(1)^{(1)}_m$ then corresponds to adding
the term ${1\over 2\pi} (dA_1 + NB_2)\wedge C_2$.}
For the finite subgroup this becomes
\be
S_{anomaly}[\mathsf{B_2},\mathsf{C_2}] = {2\pi N\over N_e N_m} \int_{M_5} \delta\mathsf{B}_2 \cup \mathsf{C}_2 \, ,
\ee
where 
\be
\mathsf{B}_2 = {N_e\over 2\pi} \, B_2 \; , \; \mathsf{C}_2 = {N_m\over 2\pi} \, C_2 \, ,
\ee
are $\mathbb{Z}_{N_e}$ and $\mathbb{Z}_{N_m}$ valued 2-cocycles, respectively.
This anomaly is clearly trivial if $N_e N_m$ is a divisor of $N$, namely if $N = r N_e N_m$ for some $r\in \mathbb{Z}$.
Gauging $\mathbb{Z}_{N_e}^{(1)} \times \mathbb{Z}_{N_m}^{(1)}$ naively gives an emergent ``dual"
$\mathbb{Z}_{N_e}^{(1)} \times \mathbb{Z}_{N_m}^{(1)}$ symmetry.
However we must take into account the mixed anomalies between 
the finite subgroups that we gauge and the  $U(1)^{(1)}$ symmetries.
In general, there is an order $N_e/\mbox{gcd}(N,N_e)$ anomaly between $\mathbb{Z}_{N_e}^{(1)}$ and $U(1)^{(1)}_m$,
with an anomaly action
\be
\label{ElectricAnomaly}
S_{anomaly}[\mathsf{B_2},C_2] =  {N\over N_e} \int \delta \mathsf{B_2} \wedge C_2 \, ,
\ee
and an order $N_m/\mbox{gcd}(N,N_m)$ anomaly between $\mathbb{Z}_{N_m}^{(1)}$ and $U(1)^{(1)}_e$,
with an anomaly action
\be
\label{MagneticAnomaly}
S_{anomaly}[B_2,\mathsf{C_2}] =  {N\over N_m} \int d B_2 \wedge \mathsf{C_2} \,  .
\ee
Since both of these anomalies are trivial for $N=rN_e N_m$, the finite part of the global symmetry is indeed
$\mathbb{Z}_{N_e}^{(1)} \times \mathbb{Z}_{N_m}^{(1)}$, and the full global symmetry is
\be
U(1)^{(1)} \times U(1)^{(1)} \times \mathbb{Z}_{N_e}^{(1)} \times \mathbb{Z}_{N_m}^{(1)}   = 
U(1)^{(1)} \times U(1)^{(1)}  \times \mathbb{Z}_{N/rk}^{(1)} \times \mathbb{Z}_{k}^{(1)}  ,
\ee
where we have defined $k \equiv N_m$.
We therefore identify the resulting theory as the $[SU(N)/\mathbb{Z}_k \times U(1)]/\mathbb{Z}_r$ theory.
Alternatively we can define $k \equiv N_e$ and obtain the 
$[SU(N)/\mathbb{Z}_{N/rk} \times U(1)]/\mathbb{Z}_r$ theory.

There are more instances in which the anomaly vanishes, leading to potentially more general gaugings.
To see them we should re-express the anomaly action in terms of the extension (Bockstein) classes
corresponding to the mixed anomalies in (\ref{ElectricAnomaly}) and (\ref{MagneticAnomaly}) \cite{Tachikawa:2017gyf}.
The electric class is associated to the short exact sequence 
\be
1 \rightarrow \mathbb{Z}_{N_e} \rightarrow U(1) \times \mathbb{Z}_{gcd(N,N_e)} \rightarrow U(1) \rightarrow 1 ,
\ee
and therefore given by 
\be
e(\mathsf{B}_2) = {\mbox{gcd}(N,N_e)\over N_e } \, \delta\mathsf{B}_2 ,
\ee
and the magnetic class is likewise given by
\be
e(\mathsf{C}_2) = {\mbox{gcd}(N,N_m)\over N_m } \, \delta\mathsf{C}_2 .
\ee
The anomaly action becomes
\be
S_{anomaly} = {2\pi N\over \mbox{gcd}(N,N_e) N_m} \int_{M_5} e(\mathsf{B}_2) \cup \mathsf{C}_2 
= \mbox{} - {2\pi N\over \mbox{gcd}(N,N_m) N_e} \int_{M_5} \mathsf{B}_2 \cup e(\mathsf{C}_2) . \nonumber \\
\ee
This is trivial if $N=rN_e N_m$, yielding the same set of possible gaugings as before.
But it is also trivial for arbitrary $N_e$ if $N_m = 1$, and for arbitrary $N_m$ if $N_e = 1$.
This agrees with the fact that the mixed anomaly allows gauging an arbitrary subgroup
of just one of the $U(1)^{(1)}$ symmetries.
More generally the anomaly is trivial if $\mbox{gcd}(N_e,N_m) = 1$.
This can be shown by a simple generalization of the $N=1$ case studied in \cite{Cordova:2023ent}.
If $\mbox{gcd}(N_e,N_m) = 1$ there exists a pair of integers $(x,y)$ such that $xN_e + yN_m = 1$. Then
\begin{align}\begin{split}
S_{anomaly} &= {2\pi N\over \mbox{gcd}(N,N_e) N_m} \int_{M_5} e(\mathsf{B}_2) \cup \mathsf{C}_2  \\
 &= {2\pi N\over \mbox{gcd}(N,N_e) N_m} \int_{M_5} (1 - y N_m) e(\mathsf{B}_2) \cup \mathsf{C}_2  \\
 &={2\pi N N_e x\over \mbox{gcd}(N,N_e) N_m} \int_{M_5} e(\mathsf{B}_2) \cup \mathsf{C}_2  \\
 &= {2\pi N x\over N_m} \int_{M_5} \delta \mathsf{B}_2 \cup \mathsf{C}_2  \\
 &= - {2\pi N x\over N_m} \int_{M_5} \mathsf{B}_2 \cup \delta \mathsf{C}_2  \\
 &=  - {2\pi Nx\over \mbox{gcd}(N,N_m)}  \int_{M_5} \mathsf{B}_2 \cup e(\mathsf{C}_2)  \\
 &= 0 \; \mbox{mod} \; 2\pi .
\end{split}\end{align}
Due to the mixed anomalies between the 
gauged discrete subgroup and the initial $U(1)^{(1)}_e \times U(1)^{(1)}_m$ symmetry, the global symmetry after gauging is
\be
U(1)^{(1)}_e \times U(1)^{(1)}_m \times \mathbb{Z}_{gcd(N,N_e)}^{(1)} \times \mathbb{Z}_{gcd(N,N_m)}^{(1)} .
\ee
Let us define $k\equiv\mbox{gcd}(N,N_m)$ and $r\equiv N/\mbox{gcd}(N,N_e N_m)$.
Since $\mbox{gcd}(N_e,N_m)=1$, we have that $\mbox{gcd}(N,N_e N_m) = \mbox{gcd}(N,N_e) \mbox{gcd}(N,N_m)$,
and then the finite part is $\mathbb{Z}_{N/rk} \times \mathbb{Z}_k$, which is the symmetry of the $[SU(N)/\mathbb{Z}_k \times U(1)]/\mathbb{Z}_r$
theory. In other words this does not lead to new theories.

\section{SymTFT for the $u(N)$ theories}
\label{symTFTU(N)}

\subsection{SymTFT and topological operators}

The $u(N)$ theories all share a $U(1)^{(1)}_e\times U(1)^{(1)}_m$ global symmetry with a mixed anomaly.
Generalizing \cite{Brennan:2024fgj,Antinucci:2024zjp}, 
we propose that the $u(N)$ SymTFT is given by
\be
\label{USymTFT}
S_{sym}^{u(N)}[B_2,C_2,h_2,f_2] = {1\over 2\pi} \int_{M_5}   [h_2\wedge dB_2 + f_2 \wedge dC_2 + NC_2\wedge dB_2] ,
\ee
where $B_2,C_2$ are $U(1)$ gauge fields and $h_2,f_2$ are $\mathbb{R}$ gauge fields.
The gauge symmetries of this theory are
\be
\delta B_2 = d\Lambda_1^B \; , \; \delta C_2 = d\Lambda_1^C \; , \; \delta f_2 = d\lambda_1^f \; , \; \delta h_2 = d\lambda_1^h .
\ee
For $N=1$, the action in (\ref{USymTFT}) reduces to the SymTFT of Maxwell theory \cite{Antinucci:2024zjp}.
The topological operators of the $u(N)$ SymTFT are given by
\be
\label{SymTFTOperators}
\begin{array}{ccc}
U_B[n_B,\Sigma_2]  = e^{in_B\oint_{\Sigma_2} B_2}\; , \;  &  U_C[n_C,\Sigma_2] = e^{in_C \oint_{\Sigma_2} C_2} \; & n_B,n_C\in \mathbb{Z} \\ [5pt]
U_h[\alpha_h,\Sigma_2] = e^{i\alpha_h\oint_{\Sigma_2} h_2}\; , \; & U_f[\alpha_f,\Sigma_2] = e^{i\alpha_f\oint_{\Sigma_2} f_2} \; & 
\alpha_h,\alpha_f \in \mathbb{R} .
\end{array}
\ee
However, these operators are not completely independent. 
The sum in the path integral over the quantized fluxes 
of $dB_2$ and $dC_2$ gives the relations
\be
\label{OperatorRelations}
U_B[Nn,\Sigma_2]U_f[-n,\Sigma_2] = 1 \; , \; U_C[Nn,\Sigma_2]U_h[n,\Sigma_2] = 1 .
\ee
So in effect we may restrict $\alpha_h,\alpha_f \in [0,1)$.
This also implies that the combinations
\be
\label{ZNOperators}
\tilde{U}_B[n_B] := U_B[n_B] U_f\left[- {n_B \over N}\right] \; , \; 
\tilde{U}_C[n_C] := U_C[n_C] U_h\left[{n_C \over N}\right] ,
\ee
are $\mathbb{Z}_N$ valued operators.
The non-trivial link-pairings are given by
\be
\langle U_B[n_B,\Sigma_2] U_h[\alpha_h,\Sigma_2']\rangle & = & e^{-2\pi i n_B\alpha_h L(\Sigma_2,\Sigma_2')} \\
\langle U_C[n_C,\Sigma_2] U_f[\alpha_f,\Sigma_2']\rangle & = & e^{-2\pi i n_C\alpha_f L(\Sigma_2,\Sigma_2')} \\
\langle U_h[\alpha_h,\Sigma_2] U_f[\alpha_f,\Sigma_2']\rangle & = & e^{2\pi i N\alpha_h \alpha_f L(\Sigma_2,\Sigma_2')},
\ee
or in terms of the operators in (\ref{ZNOperators}), by
\be
\label{LinkPairingBC}
\langle \tilde{U}_B[n_B,\Sigma_2] \tilde{U}_C[n_C,\Sigma_2']\rangle & = & e^{2\pi i {n_B n_C\over N} L(\Sigma_2,\Sigma_2')}\\
\label{LinkPairingfh}
\langle U_h[\alpha_h,\Sigma_2] U_f[\alpha_f,\Sigma_2']\rangle & = & e^{2\pi i N\alpha_h \alpha_f L(\Sigma_2,\Sigma_2')} .
\ee
The latter also follow by changing variables in the action $\tilde{B}_2 = B_2 - f_2/N$ and $\tilde{C}_2 = C_2 + h_2/N$, 
which gives\footnote{This change of variables is legal since shifting a $U(1)$ gauge field by an $\mathbb{R}$ gauge field gives a $U(1)$ gauge field.}
\be
\label{USymTFT2}
S_{sym}^{u(N)}[\tilde{B}_2,\tilde{C}_2,h_2,f_2] = {1\over 2\pi} \int_{M_5}   \left[{1\over N}h_2\wedge df_2 
+ N\tilde{C}_2\wedge d\tilde{B}_2\right] .
\ee
This formulation of the SymTFT clearly displays the $su(N) \times u(1)$ algebra, as it is just the product of the $su(N)$ and $u(1)$ SymTFT's. Depending on the objectives, one may choose to work with either formulation.

\subsection{Boundary conditions}

The 5d SymTFT corresponds to a {\em relative} 4d QFT,
and the different {\em absolute} QFT's correspond to different boundary conditions on the 5d gauge fields.
In the present case the different 4d theories will have gauge symmetries with different global structures, 
$[SU(N)/\mathbb{Z}_k \times U(1)]/\mathbb{Z}_r$, and different values of the discrete theta
parameter $\ell$.
The allowed boundary conditions are constrained by the canonical commutation relations
of the gauge fields, or equivalently
by the link-parings of the topological operators.
A consistent set of boundary conditions must fix a maximal set of mutually commuting fields at the boundary,
or equivalently a maximal set of unlinked topological operators that can end on the boundary.

A generic surface operator has the form
\be
\label{GenericSurfaceOperator}
U[n_B,n_C,\alpha_h,\alpha_f; \Sigma_2] = \tilde{U}_B[n_B;\Sigma_2] \tilde{U}_C[n_C;\Sigma_2] U_h[\alpha_h;\Sigma_2] U_f[\alpha_f;\Sigma_2] .
\ee
The condition that two such operators $U$ and $U'$ have a trivial link-pairing requires
\be
\label{TrivialLinking}
{n_B n_C' - n_C n_B'\over N}
+ N(\alpha_h \alpha_f' - \alpha_f \alpha_h') \in \mathbb{Z} .
\ee
This is equivalent to the Dirac pairing condition of the gauge theory (\ref{DiracPairing}) if we identify
\be
n_B = z_e \; , \; n_C = z_m \; , \; \alpha_f = {n_e\over N} \; , \; \alpha_h = \mbox{} - n_m .
\ee
We can therefore map the spectrum of line operators of the $[SU(N)/\mathbb{Z}_k \times U(1)]/\mathbb{Z}_r$
theory (\ref{LineOperators1}) to the surface operators of the SymTFT (\ref{GenericSurfaceOperator}) as follows
\be
\label{nB}
n_B &=& kn_1 + \ell n_2 \\
\label{nC}
n_C &=& {N\over rk} n_2 \\
\label{alphaf}
\alpha_f &=& {1\over N} (n_1 + rn_3 ) \\
\label{alphah}
\alpha_h &=& {1\over r} (n_2 - rn_4) .
\ee
This defines a maximal set of mutually unlinked surface operators,
that can therefore simultaneously end on the boundary.

The symmetry operators of the gauge theory correspond to the complementary set of surface operators that
link non-trivially with these (and possibly with each other). 
The continuous symmetries $U(1)^{(1)}_e$ and $U(1)^{(1)}_m$ are implemented by
$U_f[\alpha_f]$ and $U_h[\alpha_h]$, with $\alpha_f,\alpha_h$ unrestricted.
The operators $\tilde{U}_B[1]$ and $\tilde{U}_C[1]$ generate a finite one-form symmetry.
The relations (\ref{nB}) - (\ref{alphah}) imply that at the boundary
\be
\tilde{U}_B[k] & = & U_f\left[-{1\over N}\right] \\
\tilde{U}_C\left[{N\over r\mbox{gcd}(k,\ell)}\right] &=& \tilde{U}_B\left[-{\ell k\over\mbox{gcd}(k,\ell)}\right] 
U_h\left[-{k\over r\mbox{gcd}(k,\ell)}\right] \nonumber \\
&=& U_f\left[+{\ell\over N\mbox{gcd}(k,\ell)}\right] U_h\left[-{k\over r\mbox{gcd}(k,\ell)}\right] ,
\ee
namely that the operators on the LHS are contained in the $U(1)^{(1)}$ symmetries.
Therefore $\tilde{U}_B[1]$ generates a $\mathbb{Z}_k^{(1)}$ symmetry, and $\tilde{U}_C[1]$
generates a $\mathbb{Z}_{N/r gcd(k,\ell)}^{(1)}$ symmetry.
However these are not independent, since (\ref{nB}) - (\ref{alphah}) also imply that at the boundary
\be
\tilde{U}_B[\ell] \, \tilde{U}_C\left[{N\over rk}\right] = U_h\left[-{1\over r}\right] ,
\ee
which is also a $U(1)^{(1)}$ operator.
The finite part of the global symmetry is therefore given by
\be
\label{FiniteGlobalSymmetry2}
\left(\mathbb{Z}_{k}^{(1)} \times \mathbb{Z}_{N/r{gcd}(k,\ell)}^{(1)}\right)/\mathbb{Z}_{k/{gcd}(k,\ell)}^{(1)} 
\cong \mathbb{Z}_{N/r gcd(k,N/rk,\ell)}^{(1)} \times \mathbb{Z}_{gcd(k,N/rk,\ell)}^{(1)},
\ee
and so we reproduce the global symmetry of the boundary theory (\ref{GlobalSymmetry}).\footnote{Note that the 
LHS of (\ref{FiniteGlobalSymmetry2}) is related to (\ref{FiniteGlobalSymmetry}) by exchanging $k$ with ${N\over rk}$, 
and that the RHS is invariant under this transformation.}
The non-trivial link-pairings (\ref{LinkPairingBC}) and (\ref{LinkPairingfh}) correspond to the mixed anomalies between the two
finite symmetry factors and between the two $U(1)^{(1)}$ factors, respectively.

\section{A new SymTFT for $su(N)$ theories} 
\label{symTFTSU(N)}

The $U(N)$ theory is related to the $SU(N)$ theory by gauging a $U(1)$ global symmetry.
In the pure $SU(N)$ theory this is not a faithful symmetry, since it does not act on genuine local operators.
It does however act on the {\em baryon vertex}, which is a collection of $N$ semi-infinite Wilson lines ending at a point.
Gauging this {\em baryonic} $U(1)$ zero-form symmetry leads to both a magnetic and an electric $U(1)$ one-form symmetry.
We would like to understand this relation from the point of view of the SymTFT.

The SymTFT of the $su(N)$ theories is the $BF$ theory \cite{Witten:1998wy}
\be
\label{SUSymTFT1}
S_{su(N)}[B_2,C_2] = {N\over 2\pi} \int_{AdS_5} C_2\wedge dB_2 .
\ee
However this only incorporates the discrete global symmetry of the $su(N)$ theories.
A SymTFT that includes also the baryonic symmetry is given by
\be
\label{SUSymTFT2}
S'_{su(N)}[B_2,C_1,h_2,f_3] = {1\over 2\pi} \int_{AdS_5} \left[h_2\wedge dB_2 + f_3\wedge(dC_1 + NB_2)\right] ,
\ee
where $B_2,C_1$ are $U(1)$ gauge fields, and $h_2,f_3$ are $\mathbb{R}$ gauge fields.
Integrating out $C_1$ sets $df_3=0$ and requires $\oint f_3 \in 2\pi \mathbb{Z}$, and therefore 
implies that $f_3$ is a $U(1)$ field strength $f_3 = dC_2$.
This gives back (\ref{SUSymTFT1})
upon shifting $C_2 \rightarrow C_2 + {1\over N} h_2$.
But, as we will see, the extended theory (\ref{SUSymTFT2}) incorporates the baryonic symmetry.

The theory (\ref{SUSymTFT2}) is invariant under the gauge symmetries
\be
\delta B_2 = d\Lambda_1^B \; , \;
\delta C_1 = d\Lambda_0^C - N\Lambda_1^B \; , \;
\delta f_3 = d\lambda_2^f \; , \;
\delta h_2 = d\lambda_1^h + N\lambda_2^f .
\ee
The operators
\be 
U_f[\alpha_f,\Sigma_3] = e^{i\alpha_f\oint_{\Sigma_3} f_3} \; , \; 
U_B[n_B,\Sigma_2] =  e^{in_B \oint_{\Sigma_2} B_2 } 
\ee
with $\alpha_f\in [0,1)$ are therefore gauge invariant. 
The operators 
\be 
U_h[\alpha_h,\Sigma_2] = e^{i\alpha_h\oint_{\Sigma_2} h_2} \; , \; 
U_C[n_C,\Sigma_1] =  e^{in_C \oint_{\Sigma_1} C_1 }
\ee
are not gauge invariant in general, but can be combined with higher dimensional operators into gauge invariant combinations:
\be
\label{UhUg}
\hat{U}_h[\alpha_h,\Sigma_2,\gamma_3] &=&  U_h[\alpha_h,\Sigma_2] U_f[-N\alpha_h,\gamma_3] = e^{i\alpha_h\oint_{\Sigma_2} h_2 - iN\alpha_h\int_{\gamma_3} f_3} \\
\hat{U}_C[n_C,\Sigma_1,\gamma_2] &=&  U_C[n_C,\Sigma_1] U_B[Nn_C,\gamma_2] = e^{in_C\oint_{\Sigma_1} C_1 + iNn_C \int_{\gamma_2} B_2}
\ee
where $\partial\gamma_3 = \Sigma_2$ and $\partial\gamma_2 = \Sigma_1$.
These are {\em non-genuine} operators in the sense that they depend in general on a higher dimensional manifold.
However the operator $\hat{U}_h$ becomes a genuine surface operator 
 for $\alpha_h \in \mathbb{Z}/N$. This is understood as follows. The equation of motion for $C_1$ sets $df_3=0$, and requires
 $\oint_{\Sigma_3} f_3 \in 2\pi \mathbb{Z}$ on any closed 3-manifold $\Sigma_3$.
 This implies that for $\alpha_h \in \mathbb{Z}/N$, the operator $\hat{U}_h$ does not depend on the 3-manifold $\gamma_3$.
Furthermore since the equation of motion for $B_2$ sets $dh_2 = Nf_3$
and requires  $\oint_{\Sigma_2} h_2 - iN\int_{\gamma_3} f_3 \in 2\pi \mathbb{Z}$ on any 3-manifold $\gamma_3$,
with $\partial\gamma_3 = \Sigma_2$, it is $\mathbb{Z}_N$ valued:
\be
\left(\hat{U}_h\left[{n\over N},\Sigma_2\right]\right)^N = \hat{U}_h\left[n,\Sigma_2\right] = 1 .
\ee

The non-trivial Link-pairings are given by
\be
\langle U_f[\alpha_f,\Sigma_3] \hat{U}_C[n_C,\Sigma_1]\rangle &=& e^{2\pi i n_C \alpha_f L(\Sigma_3,\Sigma_1)} \\
\langle \hat{U}_h[\alpha_h,\Sigma_2] U_B[n_B,\Sigma_2']\rangle &=& e^{2\pi i n_B \alpha_h L(\Sigma_2,\Sigma_2')} .
\ee
The condition for trivial link-pairings between operators ending on the boundary is
\be
\label{TrivialLinking2}
n_B \alpha_h' - \alpha_h n_B' +n_C \alpha_f' - \alpha_f n_C'  \in \mathbb{Z} .
\ee
For $SU(N)$ the boundary conditions correspond to $\alpha_f = \alpha_h=0$ (or equivalently $\in\mathbb{Z}$) 
and $n_B,n_C \in \mathbb{Z}$.
In this case there is a $\mathbb{Z}_N^{(1)}$ symmetry implemented by
$\hat{U}_h\left[{n_h\over N}\right]$ acting on $U_B[n_B]$, and a $U(1)^{(0)}$ symmetry implemented by $U_f[\alpha_f]$
acting on $\hat{U}_C[n_C]$. The latter is dual to the baryon vertex of the $SU(N)$ theory.
For $PSU(N)=SU(N)/\mathbb{Z}_N$ the boundary conditions correspond to $\alpha_f = 0$,
$\alpha_h \in \mathbb{Z}/N$, $n_B \in N \mathbb{Z}$, and $n_C \in \mathbb{Z}$. In this case there is a $\mathbb{Z}_N^{(1)}$
symmetry implemented by $U_B[n_B]$ acting on $\hat{U}_h\left[{n_h\over N}\right]$, and a $U(1)^{(0)}$
symmetry implemented by $U_f[\alpha_f]$ acting on $\hat{U}_C[n_C]$, which is now the baryon vertex
of the $PSU(N)$ theory.
More general boundary conditions mix the operators $U_B[n_B]$ and $\hat{U}_h\left[{n_h\over N}\right]$,
and correspond to $[SU(N)/\mathbb{Z}_k]_\ell$.

\subsection{From $su(N)$ to $u(N)$}

Starting with the extended SymTFT of the $su(N)$ theories (\ref{SUSymTFT2}), one should be able to obtain 
the SymTFT of the $u(N)$ theories (\ref{USymTFT}) by gauging the unfaithful baryonic $U(1)$ 0-form symmetry.
Following the procedure of \cite{Antinucci:2024zjp}, we add the SymTFT of the Maxwell theory, and 
minimally-couple it to the  $su(N)$ SymTFT. The full action is given by
\begin{align}\begin{split}
\label{CoupledSymTFT}
S &=  {1\over 2\pi} \int_{M_5} \Big[ h_2\wedge dB_2 + f_3\wedge(dC_1 + NB_2) \\
&+  g_2 \wedge dG_2 + f_2 \wedge dF_2 + G_2 \wedge dF_2 + f_3\wedge G_2  \Big] ,
\end{split}\end{align}
where $G_2,F_2$ are the $U(1)$ gauge fields and $g_2,f_2$ are the $\mathbb{R}$ gauge fields 
of the Maxwell SymTFT. 
The gauge transformations of the new fields are
\be
\delta G_2 = d\Lambda_1^G \; , \; \delta F_2 = d\Lambda_1^F \; , \; \delta f_2 = d\lambda_1^f \; , \;
\delta g_2 = d\lambda_1^g + \lambda_2^f ,
\ee
and the gauge transformation of $C_1$ is modified to
\be
\delta C_1 = d\Lambda_0^C - N\Lambda_1^B - \Lambda_1^G .
\ee
This implies in particular that the $U(1)$ 0-form symmetry operator $U_{f_3}[\alpha]$
is now endable on the non-gauge-invariant surface operator $U_{g_2}[\alpha]$,
as is appropriate for a gauged symmetry.

Integrating out $C_1$ again sets $f_3 = dC_2$, where $C_2$ is a $U(1)$ gauge field.
Integrating out $G_2$ then sets $d(C_2+F_2-g_2)=0$
and the action reduces to
\be
S =  {1\over 2\pi} \int_{M_5} \left[(h_2 - Ng_2)\wedge dB_2 + f_2 dF_2 + NF_2dB_2 \right] ,
\ee
which reproduces the $u(N)$ SymTFT (\ref{USymTFT}) upon 
shifting $h_2 \rightarrow h_2 + Ng_2$,
and renaming $F_2 \rightarrow C_2$.

\section{Attempt at a holographic description}
\label{holography}

\begin{figure}[h!]
\centering
\includegraphics[scale=0.5]{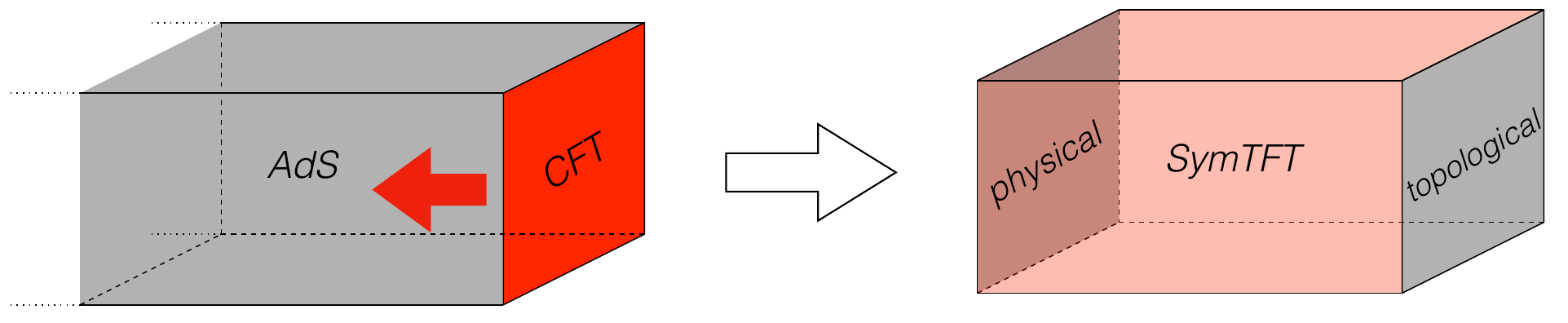}
\caption{The {\em holographic sandwich press}: the bulk of $AdS$ is squeezed into the physical boundary, and the near-boundary region of $AdS$ is expanded into the bulk of the SymTFT.}
\label{Sandwich}
\end{figure}

For theories with a holographic dual the SymTFT may be viewed as a limit of the bulk theory.
Roughly speaking, one concentrates on the bulk gauge fields dual to the boundary symmetries,
and takes the ``near boundary" IR limit (Fig.~\ref{Sandwich}).
Typically what remains is a topological field theory.

Applied to the holographic dual of ${\cal N}=4$ SYM theory, this procedure leads to the $su(N)$ SymTFT \cite{Witten:1998wy}.
The action of the NSNS and RR 2-forms $B_2, C_2$ reduces on $AdS_5\times S^5$ to
\be
\label{AdS5Action1}
S_5[B_2,C_2] =  \int_{AdS_5} \left[{1\over 2g^2} \left(|dB_2|^2 + |dC_2|^2  \right)
+{N\over 2\pi} C_2\wedge dB_2 \right] ,
\ee
where $g^2 \sim \ell_s^4/L^5$.
The kinetic terms are subdominant in the IR, and the theory reduces to 
\be
S_{IIB}^{IR}[B_2,C_2] = {N\over 2\pi} \int_{AdS_5} C_2\wedge dB_2 ,
\ee
which is the SymTFT of the $su(N)$ theories (\ref{SUSymTFT1}).

The question of how to incorporate the center-of-mass $u(1)$, the so-called singleton sector, into the holographic description
was addressed in \cite{Maldacena:2001ss}, by adding a boundary term, and in \cite{Belov:2004ht}
by carefully treating the kinetic terms.
This begs the question of whether the $u(N)$ SymTFT can be recovered from the bulk action (\ref{AdS5Action1}).

A possible strategy was presented in \cite{Bonetti:2024cjk}, where it was shown that in some cases kinetic terms may be 
reformulated in the IR as topological theories of $\mathbb{R}$ gauge fields.
For a $d$-dimensional field theory with a $U(1)$ 0-form global symmetry the $(d+1)$-dimensional
holographic bulk theory has a free massless $U(1)$ gauge field
$A_1$ with a Maxwell action:
\be
S_{d+1}[A_1] = \int_{AdS_{d+1}} {1\over 2g^2} dA_1 \wedge *dA_1 
= \int_{AdS_{d+1}} {1\over 2g^2} {L^{d-3}\over z^{d-3}} dA_1 \wedge \tilde{*} dA_1,
\ee
where $\tilde{*}$ is the flat space Hodge-star. 
This can be reformulated by introducing a Lagrange multiplier field $f_{d-1}$ with action
\be
S_{d+1}[A_1,f_{d-1}] = \frac{1}{2\pi}\int_{AdS_{d+1}} \left[f_{d-1}\wedge dA_1 - {g^2\over 4\pi}{z^{d-3}\over L^{d-3}} f_{d-1}\wedge \tilde{*}f_{d-1}\right]  .
\ee
Integrating out $f_{d-1}$ sets $f_{d-1} = {2\pi\over g^2} {L^{d-3}\over z^{d-3}} \tilde{*}dA_1$, and we recover the original action.
On the other hand in the near-boundary IR limit $z \rightarrow 0$ the second term vanishes, under the condition that $d > 3$, and
we are left with a topological theory
\be
\label{U1SymTFT}
S^{IR}_{d+1}[A_1,f_{d-1}] = \frac{1}{2\pi} \int_{AdS_{d+1}} f_{d-1}\wedge dA_1   .
\ee
This has an emergent gauge symmetry
 \be
 f_{d-1} \rightarrow f_{d-1} + d\lambda_{d-2} ,
 \ee
 where $\lambda_{d-2}$ is globally defined, so the field $f_{d-1}$ may be regarded as an $\mathbb{R}$ gauge field.
The theory (\ref{U1SymTFT}) is therefore the SymTFT of a $d$-dimensional theory with a $U(1)$ 0-form global symmetry \cite{Brennan:2024fgj,Antinucci:2024zjp}. 

Applying this idea to (\ref{AdS5Action1}), we introduce 
two Lagrange multiplier fields $h_2,f_2$, and reformulate the bulk theory as
\begin{align}\begin{split}
\label{AdS5Action2}
S_5[B_2,C_2,h_2,f_2] &= {1\over 2\pi} \int_{AdS_5}  
\Big[ h_2 \wedge dB_2 + f_2 \wedge dC_2 + NC_2\wedge dB_2  \\
 &- {g^2\over 4\pi} {L\over z} 
 (h_2 \wedge \tilde{*} h_2 + f_2 \wedge \tilde{*} f_2) \Big].
\end{split}\end{align}
Integrating out $h_2, f_2$ sets 
\be
\label{EMdual}
h_2  = {2\pi\over g^2}{z\over L} \tilde{*} dB_2 
\; , \; 
f_2 = {2\pi\over g^2} {z\over L} \tilde{*} dC_2 ,
\ee
and we recover the original action.
The topological part of the new action is precisely the $u(N)$ SymTFT (\ref{USymTFT}),
but crucially, in this case we cannot neglect the non-topological terms in the IR. 
The near-boundary limit of the supergravity action does not reduce to the SymTFT in this case.

\subsection{Branes and topological operators} 

There is a related difficulty in trying to identify the topological operators of the SymTFT in terms of branes in 
Type IIB string theory.
The surface operators $U_{B}[n_B]$ and $U_{C}[n_C]$ are easily identified with fundamental strings and D1-branes, respectively.

Following our proposal in \cite{Bergman:2024aly} (see also \cite{Calvo:2025kjh,Calvo:2025usj}) one is tempted to identify the continuous surface operators 
$U_{h}[\alpha_{h_2}]$, $U_{f}[\alpha_{f_2}]$ 
with the non-BPS D6-brane and non-BPS NS6-brane wrapping $S^5$.\footnote{The existence of non-BPS NS-branes is implied by 
S-duality \cite{Houart:1999rf}.}
This almost works, but fails in an interesting way.
As argued in \cite{Bergman:2024aly}, the worldvolume action of the non-BPS D6-brane in the tachyon vacuum has a
remnant given by 
\begin{align}\begin{split}
\label{D6Action}
S^{\text{vac}}_{\widetilde{\text{D6}}}  &= \alpha \int_{S^5\times \Sigma_2} \left(dC_6 
- {1\over 2\pi} dC_4 \wedge (B_2 + F_2^{wv}) + \cdots \right) \\
&= \alpha \int_{\Sigma_{2}} \left({1\over g^2} *dC_2 - {N\over 2\pi}(B_2 + F_2^{wv})\right) ,
\end{split}\end{align}
where  $\alpha$ is a continuous parameter taking values in $[0,2\pi)$, and $F_2^{wv}$ is the field strength of the worldvolume gauge field.\footnote{We can redefine $\alpha \in [0,1)$ by absorbing the $2\pi$ 
from the second term and redefining $g^2$.}
The appearance of the combination $B_2 + F_2^{wv}$ is required by gauge invariance under gauge transformations
of $B_2$, as in the case of BPS D-branes.\footnote{See also \cite{Calvo:2025usj}.}
Crucially, in the second equality we took into account the RR flux on $S^5$, which leads to the second term 
in the worldvolume action.
In the absence of this term we would identify the non-BPS D6-brane with the operator $U_f[\alpha]$.
However the presence of this term requires $\alpha$ to be quantized.
The worldvolume path integral includes a sum over worldvolume magnetic fluxes on $\Sigma_2$,
which vanishes unless $\alpha \in 2\pi \mathbb{Z}/N$.\footnote{A similar mechanism was observed
in the context of the axial symmetry of QED in \cite{Karasik:2022kkq,GarciaEtxebarria:2022jky}.}
So in fact the non-BPS D6-brane, with $\alpha = 2\pi n/N$, corresponds to the $\mathbb{Z}_N$-valued
operator $\tilde{U}_B[n] = U_B[n] U_f[-n/N]$.
A similar conclusion holds for the non-BPS NS6-brane, which corresponds to the $\mathbb{Z}_N$-valued operator
 $\tilde{U}_C[n] = U_C[n] U_h[n/N]$.
This is consistent with the fact that the dual theory is an $su(N)$, rather that $u(N)$, theory.
In fact if we keep track of the metric factors, the first term in (\ref{D6Action}) is subdominant to the second term in the near
boundary limit, and the non-BPS D6-brane reduces to a collection of fundamental strings.
Similarly the non-BPS NS6-brane reduces at the boundary to a collection of D1-branes.

Going into the bulk, the non-BPS 6-branes correspond to the symmetry operators of the full bulk theory (\ref{AdS5Action1}).
The equations of motion
\be
d\left({1\over g^2} *dB_2 +{N\over 2\pi} C_2\right) &= & 0\\
d\left({1\over g^2} *dC_2 -{N\over 2\pi} B_2\right) &= & 0 ,
\ee
imply that the quantities in parentheses can be interpreted as the Hodge-duals of conserved $U(1)$ 3-form currents.
In other words the operators
\be
U_f[\alpha_f,\Sigma_2] = e^{i\alpha_f\oint_{\Sigma_2} \left({1\over g^2} *dC_2 -{N\over 2\pi} B_2\right)}  \; , \;
U_h[\alpha_h,\Sigma_2] = e^{i\alpha_h\oint_{\Sigma_2} \left({1\over g^2} *dB_2 +{N\over 2\pi} C_2\right)} \, ,
\ee
are topological. But they are not invariant under large gauge transformations of $B_2, C_2$, unless
$\alpha_f, \alpha_h \in \mathbb{Z}/N$. This is precisely the quantization condition that the non-BPS 6-brane
realization implies.

\subsection{General condition for holographic SymTFT of a $U(1)$ symmetry} 

Generalizing to a $U(1)$ $(p+1)$-form gauge field in $AdS_{d+1}$ we will get a condition 
for realizing the SymTFT of a $U(1)$ $p$-form symmetry in the near boundary limit.
We reformulate the Maxwell action
\begin{align}\begin{split}
S_{d+1}[A_{p+1}] &= \int_{AdS_{d+1}} {1\over 2g^2} dA_{p+1} \wedge *dA_{p+1}  \\
&= \int_{AdS_{d+1}} {1\over 2g^2} {L^{d-2p-3}\over z^{d-2p-3}} dA_{p+1} \wedge \tilde{*} dA_{p+1},
\end{split}\end{align}
by introducing a Lagrange multiplier field $f_{d-p-1}$ as
\be
\label{HolographicSymTh1}
S_{d+1}[A_{p+1},f_{d-p-1}] = \frac{1}{2\pi} \int_{AdS_{d+1}} \left[f_{d-p-1}\wedge dA_{p+1} - 
{g^2\over 4\pi}{z^{d-2p-3}\over L^{d-2p-3}} f_{d-p-1}\wedge \tilde{*}f_{d-p-1}\right]  . \nonumber \\
\ee
This reduces in the $z\rightarrow 0$ limit to the SymTFT of a $U(1)$ $p$-form symmetry:
\be
\label{psymTFT}
S_{d+1}^{IR}[A_{p+1},f_{d-p-1}] = \frac{1}{2\pi} \int_{AdS_{d+1}} f_{d-p-1}\wedge dA_{p+1} ,
\ee
provided that $d > 2p + 3$.
Alternatively, we can reformulate the above Maxwell theory using electromagnetic duality:
\begin{align}\begin{split}
S_{d+1}[A_{d-p-2}] &= \int_{AdS_{d+1}} {1\over 2\tilde{g}^2} dA_{d-p-2} \wedge *dA_{d-p-2}  \\
&= \int_{AdS_{d+1}} {1\over 2\tilde{g}^2} {z^{d-2p-3}\over L^{d-2p-3}} dA_{d-p-2} \wedge \tilde{*} dA_{d-p-2},
\end{split}\end{align}
and then introduce a Lagrange multiplier field $f_{p+2}$:
\be
\label{HolographicSymTh2}
S_{d+1}[A_{d-p-2},f_{p+2}] = \int_{AdS_{d+1}} \left[\frac{1}{2\pi} f_{p+2} \wedge dA_{d-p-2} +
{1\over 2\tilde{g}^2} {L^{d-2p-3}\over z^{d-2p-3}} f_{p+2}\wedge \tilde{*}f_{p+2}\right] .
\ee
In this case the bulk theory reduces in the $z\rightarrow 0$ limit to the SymTFT for the dual magnetic $U(1)$ $(d-p-3)$-form symmetry,
\be 
\label{d-p-3symTFT}
S_{d+1}^{IR}[A_{d-p-2},f_{p+2}] =\frac{1}{2\pi} \int_{AdS_{d+1}}  f_{p+2} \wedge dA_{d-p-2},
\ee
provided that $d<2p+3$.
Note that this agrees with the condition for the SymTFT of a $U(1)$ $p$-form symmetry by replacing
$p\rightarrow d-p-3$.

\section{Conclusions}\label{Conclusions}

In this note, we have discussed the SymTFT for Yang-Mills theories with gauge algebra $u(N) = su(N) \times u(1)$. The variant is specified by a choice of gauge group $[SU(N)/\mathbb{Z}_k \times U(1)]/\mathbb{Z}_r$ and a  $\mathbb{Z}_{rk}$-valued 
discrete theta parameter $\ell$. We studied the gapped boundary conditions of the SymTFT, reproducing the expected global variants, their symmetries, and their anomalies. The $u(N)$ Yang-Mills theory is obtained by gauging an unfaithful baryonic $U(1)$ symmetry in $su(N)$ Yang-Mills theory. This procedure can be nicely captured by the SymTFT language once the unfaithful symmetry is included into the $su(N)$ SymTFT. 

We have also discussed the embedding of the $u(N)$ SymTFT in holography. Perhaps contrary to naive expectation, but in full agreement with \cite{Witten:1998wy}, there is no near-boundary limit of the bulk supergravity action that yields the $u(N)$ SymTFT. One always recovers the $su(N)$ SymTFT. Interestingly, non-BPS D-branes, which we have proposed to describe symmetry defects in holography, capture this effect and generate the symmetry of the $su(N)$ theories in the boundary. Finally, we have shown that this failure is more general, and the SymTFT for a $U(1)$ $p$-form symmetry in $d$-dimensions will arise from holography only if $d>2p+3$. 

This is perhaps related to the issue of allowed boundary conditions in $AdS$ holography.
For $d>2p+3$ the only (Lorentz and conformal invariant) boundary condition allowed for a $(p+1)$-form gauge field $A_{p+1}$
is Dirichlet ({\em aka standard}). In this case the boundary theory has a global $U(1)$ $p$-form symmetry,
and the corresponding SymTFT (\ref{psymTFT}) is obtained in the near boundary limit of the bulk action in $AdS$.
Outside of this range, namely for $d\leq 2p+3$, a Neumann ({\em aka alternative}) boundary condition is also allowed 
\cite{Witten:2003ya,Marolf:2006nd,Esmaeili:2021szb}.
This corresponds to gauging the global $U(1)$ $p$-form symmetry in the boundary theory,
which in turn yields a magnetic $U(1)$ $(d-p-3)$-form symmetry.
However the new boundary theory is not incorporated in the SymTFT of (\ref{psymTFT}).
This is consistent with the condition $d>2p+3$ for this SymTFT to be holographic.

Despite this finding one may speculate that 
perhaps there is a more general way of thinking about symmetry theories in holographic settings.
In particular, just dropping the non-topological terms in (\ref{AdS5Action2}) gives the $u(N)$ SymTFT (\ref{USymTFT}).
If instead we first dualize $C_2$ to $C_1$ and then introduce Lagrange multiplier fields $h_2, f_3$:
\begin{align}\begin{split}
\label{AdS5Action3}
S_5[B_2,C_1,h_2,f_3] &= {1\over 2\pi} \int_{M_5} \Big[h_2 \wedge dB_2 - f_3\wedge (dC_1 + NB_2)   \\
& -  {g^2\over 4\pi} {L\over z} h_2\wedge \tilde{*}h_2 + {\pi\over 2 g^2} {z\over L} f_3 \wedge \tilde{*}f_3 \Big] ,
\end{split}\end{align}
the topological part by itself gives the extended $su(N)$ SymTFT (\ref{SUSymTFT2}), although we cannot ignore 
all the non-topological terms in the near-boundary limit.
It may be worth exploring to what extent the connection of SymTFT to holography can be extended beyond
the strict near-boundary limit.

%===============================================================================
\subsection*{Acknowledgements}
%===============================================================================
%===============================================================================

We thank Daniel Brennan and I\~naki Garc\'ia Etxebarria for discussions. O.B. is supported in part by the Israel Science Foundation under grant No. 1254/22,
and by the US-Israel Binational Science Foundation under grant No. 2022100. F.M. and D.R.-G are partially supported by the Spanish
national grant MCIU-22-PID2021-123021NB-I00.  E.G.-V. is supported by MIUR PRIN Grant 2020KR4KN2 “String Theory as a bridge between Gauge Theories and Quantum Gravity” and by INFN Iniziativa Specifica ST\&FI.
\addcontentsline{toc}{section}{Bibliography}

\bibliography{nonBPSbib}
\bibliographystyle{JHEP}

\end{document}